\documentstyle[epsfig]{mn}
\begin{document}
\renewcommand\floatpagefraction{.99}

\title[]
{
On the peak radio and X-ray emission from neutron star and black hole
candidate X-ray transients
}
\author[R.P. Fender \& E. Kuulkers] 
{R.P. Fender$^{1}$\thanks{email : rpf@astro.uva.nl},
 E. Kuulkers$^{2,3}$ 
\\
$^{1}$ Astronomical Institute `Anton Pannekoek' and Center for High Energy
Astrophysics, University of Amsterdam, Kruislaan 403, \\
1098 SJ Amsterdam, The Netherlands. \\
$^{2}$ Space Research Organization Netherlands,
Sorbonnelaan 2, 3584 CA Utrecht, The Netherlands.\\
$^{3}$ Astronomical Institute, Utrecht University, P.O. Box 80000,
3507 TA Utrecht, The Netherlands\\}

\maketitle

\begin{abstract}

We have compiled and analysed reports from the literature of
(quasi-)simultaneous observations of X-ray transients at radio and
X-ray wavelengths and compared them with each other and with more
unusual radio-bright sources such as Cygnus X-3, GRS 1915+105 and
Circinus X-1. There exists a significant ($>97$\% likelihood)
positive (rank) correlation between the peak X-ray flux P$_X$ and radio flux
density P$_R$ for the black hole candidate (BHC) systems, and a
marginally significant positive (rank) correlation
for the neutron star (NS) systems.  This is further evidence for a
coupling between accretion and outflows in X-ray binary systems, in
this case implying a relation between peak disc-accretion-rate and the
number of synchroton-emitting electrons ejected. However, we also show
that the distribution of `radio loudness', P$_R$/P$_X$, is
significantly different for the two samples, in the sense that the
BHCs generally have a higher ratio of P$_R$/P$_X$.  The origin of this
discrepancy is uncertain, but probably reflects differences in the
energetics and/or radiative efficiency of flows around the neutron
stars and black holes; we briefly discuss some of these possibilities.
Furthermore, the data for the two recently discovered `fast
transients' (FTs), XTE J0421+560/CI Cam and SAX J1819.3-2525/V4641 Sgr
are entirely compatible with the distribution of BHCs.  As at least
three of the BHCs and both FTs have been directly resolved into mildly
relativistic jets, it seems likely that such outflows are the origin
of the radio emission in all BHC and FT transients, and probably also
for the NS transients.  We further note that range of X-ray and radio
fluxes observed from the unusual superluminal source GRS 1915+105 are
also entirely compatible with the distribution for transients,
implying that there is nothing special about the physics of jet
formation in that system.  We conclude that these data point to the
formation of a mildly relativistic jet whose luminosity is a function
of the accretion rate, in the majority, if not all, of X-ray transient
outbursts, but whose relation to the observed X-ray emission is
dependent on the nature of the accreting compact object.

\end{abstract}



\section{Introduction}

\begin{figure}
\leavevmode\epsfig{file=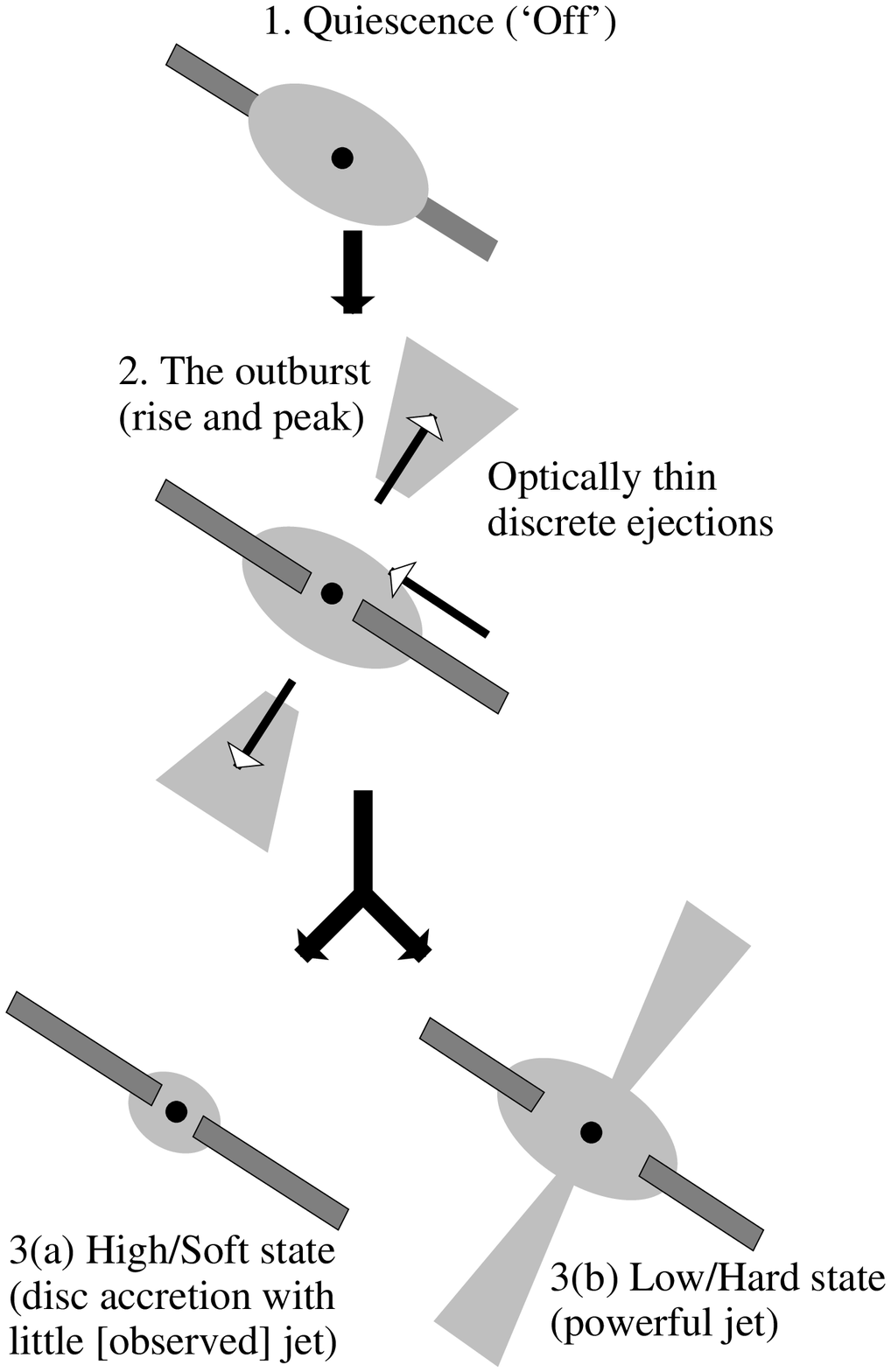,width=7.7cm,angle=0,clip}
\caption{A sketch of the phenomenology of BHC transient
outbursts. Systems transition rapidly from quiescence, 
to a brighter state, via a period in which the flux density rises
rapidly. The brighter state is generally either disc-dominated
(the high/soft state) or power-law dominated (the low/hard state) in
X-rays; our physical interpretation of these states is represented in
the figure. The reasons for the bifurcation between stages 2 and 3 is
not clear, although it seems the systems arriving at 3b may be
accreting at lower rates. In this work it is the discrete ejection
events generated at stage 2 and observed at radio wavelengths, which
we are comparing against the peak X-ray flux.
}
\end{figure}

\begin{table*}
\caption{Soft X-ray and radio peak fluxes for neutron star and
black hole X-ray transients. Peak X-ray flux, P$_X$ is
measured in soft ($\leq 12$ keV)
band; Peak radio flux density,  P$_R$,  is at 5 GHz (6 cm) or corrected from nearest
measurement using a spectral index of $\alpha=-0.5$.
Abbreviations : NS - Neutron Star; BHC - Black Hole Candidate; FT -
Fast transient}
\centering
\begin{tabular}{rrccccccl}
\hline
Name & Other & Class & Outburst & P$_X$  & P$_R$ & P$_R$/P$_X$ & Resolved & Refs\\
     & name  && date     & (Crab)&  (mJy) & (mJy/Crab)& Jet ? & \\
\hline
1A 1742-289 && NS?& 1975     & 2.0 & 215 &110& N&B76,E75,D76\\
Cen X-4     &V822 Cen& NS & 1979 & 5  & 8.0 &1.6& N & K80,H79,H88\\
Aql X-1&4U 1908+00  &NS& 1990$^*$ & 0.3--1.6 & 0.5&0.3--1.7 &N& H90,H95,C97,ASM\\
4U 1730-335 & {\bf The} Rapid & NS & 1996 Nov& 0.38 & 0.5 &1.3& N&M00,G99\\
            & Burster&& 1997 Jun& 0.36 & 0.21&0.58 & N&M00,G99\\
            &&& 1998 Jan& 0.27 & $<$0.39&$<1.4$&&M00, G99\\
SAXJ1808.4-3658 &{\bf The} msXRP &NS& 1998 & 0.08 & 0.8&10.0 &N& G98,GSG99\\
XTE J1806-246  &2S1803-245& NS& 1998 & 0.7 & 0.8&1.1&N& HRM,R99\\
\hline
1A 0620-00 & V616 Mon  &BHC & 1975 & 45 & 190 &4.2&Y?& H95,C97,K99\\
GS 2000+25 & QZ Vul &BHC& 1988& 12 & 6&0.5 &N& H95,C97,K99\\
GS 2023+338 & V404 Cyg & BHC& 1989 & 21 & 1800&86 &N& H95,H92,C97\\
GS 1124-68 & GU Mus &BHC& 1991& 8 & 140&18 &N& H95,C97,K99\\
GRO J0422+32 & V518 Per & BHC&1992 & 3 & 8&2.7 &N& S94\\
GRO J1655-40 &V1033 Sco& BHC & 1994 Aug &1.7& 3700&2200 &Y&C97,T96\\
             &&& 1995 Aug &1.3 &$<0.5$ &$<0.4$&&C97,T96\\
	     &&& 1996 May &3.2 &25 &7.8& N &H97,ASM\\ 
GRS 1716-249 & GRO J1719-24 & BHC & 1995 & 0.43 & 15.5 & 36 &N& H96 \\
GRS 1739-278   && BHC& 1996 & 1.0 & 10.0 &10.0&N& HRM,ASM \\
XTE J1755-324  && BHC& 1997 & 0.19 & $<0.3$&$<1.6$ && O97,ASM\\
GS 1354-64 & BW Cir & BHC& 1997 & 0.07 & 2.8&40 &N& B00\\
4U 1630-47 & GX 337+0 & BHC& 1998  & 0.43 & 2.6&6.0 &N& H99\\
XTE J2012+381  && BHC& 1998 & 0.22 & 3.2&15 &N& HRM,V00\\
XTE J1550-564 && BHC& 1998 & 6.6 & 160&24 &Y& W00,H01,ASM\\
XTE J1748-288 && BHC& 1998 & 0.51 & 440&860 &Y& HRM,ASM\\
XTE J1859+226 && BHC& 1999 & 1.08 & 70&65 &N& HRM,ASM \\
XTE J1118+480 && BHC& 2000 & 0.05 & 6&120 &N& GGP,HRM,K00 \\
\hline
XTE J0421+560 &CI Cam & FT & 1998 & 1.9 & 750&395 &Y& B99,C00\\ 
SAX J1819.3-2525 &V4641 Sgr & FT/BHC &1999 & 12.2 & 400&33 &Y& H00\\
Cyg X-3 & 4U 2030+40 & ? & 1972-2000 & 0.05--0.55 & 180--200000&20-20000 &Y& GGP,ASM\\
GRS 1915+105 & & BHC & 1994-2000 & 0.25--2.30 & 1--250&0.4--1000 &Y& P97\\
Cir X-1 & 4U 1516-56 & NS & 1970s & 0.5--1.5 & 1000--5000&665--10000 &N & K76,H78\\
        &           &    & 1990s & 2.0--3.5 & 40 &11--20&Y & F97 \\
\hline
\end{tabular}
\flushleft
$^*$ -- no simultaneous X-ray coverage (only optical) -- range in X-rays
estimated from several other outbursts. 
{\bf References :} 
E75 = Eyles et al. (1975);
B76 = Branduardi et al. (1976);
D76 = Davis et al. (1976);
K76 = Kaluzienski et al. (1976);
H78 = Haynes et al. (1978);
H79 = Hjellming (1979);
K80 = Kaluzienski, Holt \& Swank (1980);
H88 = Hjellming et al. (1988);
H90 = Hjellming et al. (1990);
H92 = Han \& Hjellming (1992);
S94 = Shrader et al. (1994);
H95 = Hjellming \& Han (1995);
H96 = Hjellming et al. (1996);
T96 = Tavani et al. (1996);
C97 = Chen, Shrader \& Livio (1997);
F97 = Fender (1997);
H97 = Hunstead, Wu \& Campbell-Wilson (1997);
O97 = Ogley, Ash \& Fender (1997);
P97 = Pooley \& Fender 1997;
G98 = Gilfanov et al. (1998);
B99 = Belloni et al. (1999);
G99 = Guerriero et al. (1999);
GSG99 = Gaensler, Stappers \& Getts (1999);
H99 = Hjellming et al. (1999);
K99 = Kuulkers et al. (1999);
R99 = Revnivtsev, Borozdin \& Emelyanov (1999); 
B00 = Brocksopp et al. (2000);
C00 = Clark et al. (2000);
H00 = Hjellming et al. (2000);
H01 = Hannikainen et al. (2001);
K00 = Kuulkers (2000);
M00 = Moore et al. (2000) ;
W00 = Wu et al. (2000);
V00 = Vasiliev, Trudolyubov \& Revnivtsev (2000);
HRM = Hjellming, Rupen \& Mioduszewski (private communication); 
GGP = Pooley (private communication); 
ASM = public XTE ASM data via {\bf xte.mit.edu}
\end{table*}

Radio emission from X-ray binaries is generally believed to originate
in incoherent synchrotron emission from a large volume of electrons
(and maybe also positrons) spiralling around magnetic field lines and
being ejected with large bulk velocities
from the binary system. When resolved, these ejecta are
often observed to be composed of apparently discrete clumps moving along well
collimated (opening angle $\leq 20^{\circ}$) trajectories at highly
relativistic ($v/c$ probably $\geq 0.9$ in at least four cases).
For recent reviews see Hjellming \& Han (1995), Mirabel \& Rodriguez
(1999) and Fender (2000a,b).

The strongest radio emission observed from X-ray binaries is that
associated with major flares, commonly observed during X-ray
outbursts of transient sources (Hjellming \& Han 1995; Kuulkers et
al. 1999). These transient systems are believed to undergo such
outbursts due to disc instabilities, and may contain neutron stars
(Campana et al. 1998) or black holes (Tanaka \& Lewin 1995;
Tanaka \& Shibazaki 1996; Charles 1998). 

Increasingly, strong connections between the accretion process
(observed in X-rays), and the jet/ejection process (observed at radio
wavelengths) are becoming apparent. In the persistent neutron star Z
sources a relation between position on the `Z' (representing the X-ray
colours) and radio emission has been established (Penninx et al. 1988;
Hjellming \& Han 1995 and references therein). In the persistent black
hole candidates Cyg X-1 and GX 339-4 a very clear correlation between
radio and soft X-ray emission has been established when the sources
are in the Low/Hard X-ray state (Brocksopp et al. 1999 and references
therein, Corbel et al. 2000 and references therein), switching to an
anticorrelation in the High/Soft X-ray state (Fender et al. 1999b).  In
addition, the brightest persistent black hole and neutron star X-ray
binaries have been found to have an approximately common radio
luminosity (Fender \& Hendry 2000).

In Fig 1 we summarise our understanding of the relation of radio
emission to transient BHC systems. From quiescence (phase 1 in Fig 1),
in which systems generally spend 99\% of their time, they are observed
to evolve rapidly within a day or so to a much brighter state,
presumably indicative of an increased accretion rate (e.g. Chen et
al. 1997). Subsequently the systems remain bright and only slowly fade
away, often taking months to return to their pre-outburst levels. The
X-ray spectrum in this stage is generally either disc- or power-law
dominated, corresponding to the canonical high/soft or low/hard X-ray
states respectively. These two states have dramatically different
radio properties (Fender 2000a,c; Fender et al. 1999b) from which we
infer differing accretion/outflow couplings, as indicated by phases 3a
and 3b in Fig 1 respectively.  However this is not directly relevant
to this study; here we focus on the discrete ejection events which
appear to be associated with the transition phase (stage 2 in Fig 1)
at the start of the outburst.  For the NS systems, which are all
(probably) essentially similar (ie. low magnetic field `atoll'
sources) there is no clear distinction in post-peak X-ray behaviour,
and as with the BHCs the optically thin radio emission observed during
transient outbursts is probably associated with the rapid ejection of
matter during the rise to peak of the outburst (stage 2 in Fig 1).

Whilst in the early years of X-ray astronomy few radio counterparts to
transient systems were reported, in the past decade many more such
systems have been found, in part due to a greater awareness of the
ubiquity and significance of the radio emission.  In this paper we
gather together all the data for BHC and NS transients for which
there are reported X-ray and radio observations, with the aim of
establishing whether there is a clear relation between X-ray and radio
emission from neutron star and black hole candidate X-ray transients
both within each class and across the classes.

\begin{figure*}
\leavevmode\epsfig{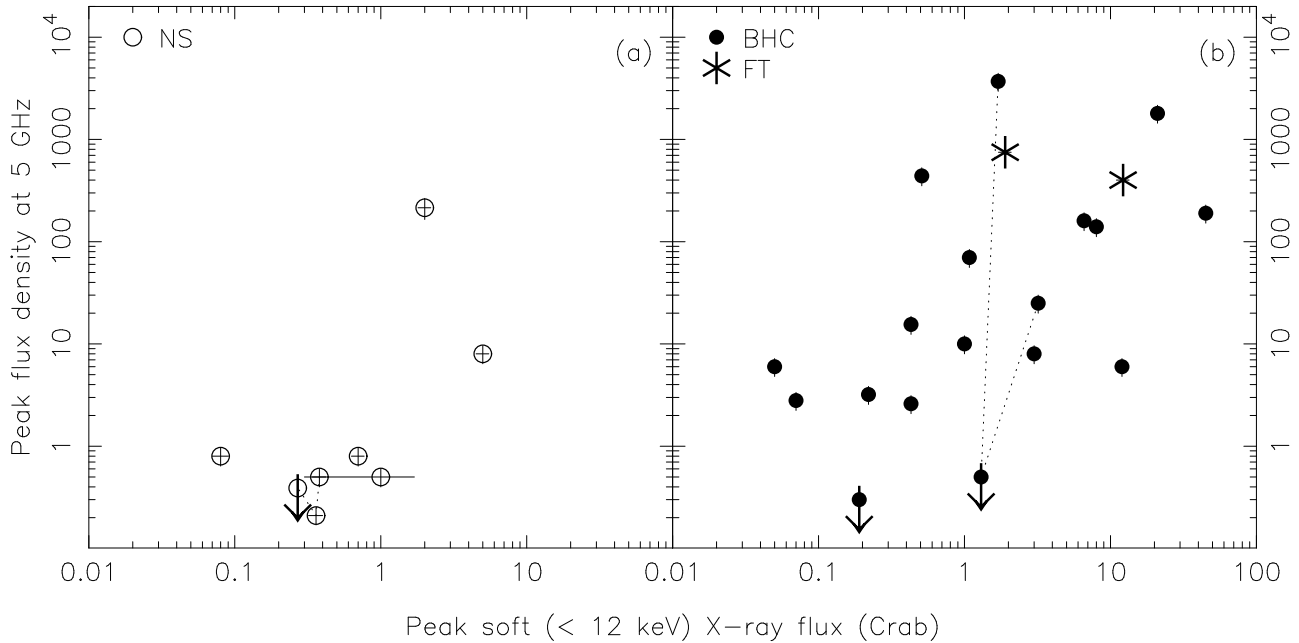}
\caption{Distribution of peak radio flux density P$_R$ as a function
of peak soft X-ray flux P$_X$ for neutron star (a) and black hole
candidate (b) X-ray transients, from the data in table 1.  The
three connected points in the NS figure indicate the two detections
and one upper limit for 4U 1730-335 (The Rapid Burster).  Also
included with the BHCs are the `fast-transient' systems. The connected
points and upper limit on the BHC figure indicates the dramatic switch
from radio-loud to radio-quiet and back to radio-loud again by GRO
J1655-40 in 1994 Aug, 1995 Aug and 1996 May.}
\end{figure*}

\section{The data}

Table 1 lists the available information for the 22 X-ray transients
for which we have some information on the peak X-ray and radio flux,
subdivided into 6 neutron-star systems (NSs) and 16 black-hole
candidates (BHCs).  In addition we tabulate the same information for
the two recently discovered `fast transients' (FTs), XTE J0421+560/CI
Cam and SAX J1819.3-2525/V4641 Sgr (the latter of which is likely to
be a BHC -- Orosz et al. 2000a, b), and several unusual sources of
recurrent X-ray and radio outbursts, namely Cyg X-3, GRS 1915+105 and
Cir X-1. As well as information on the source name, outburst date,
peak X-ray and radio fluxes and references, we indicate whether the
radio emission from the source has been directly resolved by radio
interferometry. This reveals that none of the 7 NS transient events,
three (maybe four)
of the 16 BHC events, and all of the FTs and unusual recurrent systems
have been directly resolved into jets.  The data sets for the NS and
BHC plus FT systems are plotted separately in Figs 2(a) and 1(b)
respectively.

We briefly discuss below how the peak X-ray and radio fluxes presented
in Table 1 were estimated and compiled.

\subsection{Radio}

Radio data are based on peak observed flux density at a frequency of 5
GHz. Where measurements at 5 GHz are not available, we assume a
spectral index of $\alpha = \Delta \log S_{\nu} / \Delta \log \nu =
-0.5$ in order to estimate the flux density at 5 GHz based on other
observations. While most transient outbursts are accompanied by
optically thin events for which this will be a good approximation,
some systems produce flat or inverted (in the sense that $\alpha \leq
0$) radio spectra, which are generally associated with low/hard X-ray
states (Fender 2000c). However, in nearly all cases there is an
initial optically thin event before the radio emission inverts and
becomes dominated by the flat-spectrum component, and so a comparison
with transients which achieve the high/soft state is probably
warranted (see Fig 1; an exception may be XTE J1118+480 which does not
appear at any stage to have shown an optically thin radio spectrum --
G. G. Pooley \& R. M. Hjellming, private communication).

In nearly all cases radio coverage is far less complete
than X-ray coverage, and so we may often miss the peak. In particular,
in nearly all cases the outburst is observed first in X-rays and then
followed-up at radio and other wavelengths.  However, since there is
no obvious way of compensating for what is essentially an unknown
effect, we can only assume that this affects all sources in a more or
less similar way. In an attempt to compensate for this, we somewhat
arbitrarily add a 20\% uncertainty to all our radio measurements
when representing them in the figures or statistical tests.

\subsection{X-ray}

Most X-ray transients are discovered by all-sky monitors early on
during the rise of the outburst, so that their peaks are well covered.
Such all-sky monitors operate nominally in the 1-10 keV energy range,
although the range covered is not always the same (see Chen et
al. 1997). The most notable exceptions are those transients which
were observed during 1992-1995; at that time the only all-sky monitors
were CGRO/BATSE (20-300 keV) and WATCH/GRANAT (8-20 keV; only active
in 1992--1993). Fortunately this applies in our case only to GRO
J0422+32, the 1994 and 1995 outbursts of 
GRO J1655-40, and GRS 1716-249. About three quarters
of the transients
have been discovered or monitored in the last 5 years, during
which the RXTE/ASM was active.

Since spectral data are not always available in the literature, we
decided to convert the X-ray intensities to units of the Crab count
rate in the same energy band. Note that in this case we do not take
into account interstellar extinction, which may alter the intrinsic
source strength.  Such changes are most severe in the Galactic center
region. However, we estimate that this changes the peak X-ray fluxes
only by a factor between $\sim$2--5, depending on the exact spectral
shape at the time of the peak and the interstellar absorption column.

In order to take into account the energy range covered by the
instrument in question and the effect of interstellar extinction we,
again somewhat arbitrarily, assume an error in the peak flux of 10\%
for use in figures and statistical tests.

A note is warranted regarding 1A 1742-289, whoses NS credentials are
not 100\% certain.  Lewin et al. (1976) derived source positions for
three burst sources in the Galactic centre region. For one of them,
MXB 1743-29, two equally likely positions separated by $\sim 0.3$
degrees were found. It was noted that A1742-289 lies within the
(overlapping) error circles, and that possibly A1742-289 is the burst
source MXB 1743-29. Subsequently there was the detection of a bursting
and eclipsing X-ray source by ASCA consistent with the position of 1A
1742-289 (Maeda et al. 1996).  However, Kennea \& Skinner (1996) found
that the source found by ASCA cannot be the quiescent counterpart of
1A 1742-289, based on the absence of eclipses during the outburst in
1975.  Since 1A 1742-289 lies just next to the Galactic Center the
chances are significant that they are indeed not the same source and
therefore the nature of the compact object in 1A 1742-289 remains an
open question. This is a salient point as, Cir X-1 aside, this is the
brightest radio source associated with a neutron-star X-ray binary.

\subsection{Physical meaning of the peak fluxes}

In comparing the X-ray and radio peak fluxes, what are we physically
investigating ? In X-rays, the peak flux is generally supposed to
represent the instantaneous dissipation rate of accretion energy, and
as such is proportional to the (disc) mass inflow rate, i.e.  P$_X
\propto \dot{m}_{\rm disc}$. The radio emission however is produced by
a plasma in which dissipation of internal energy is not instant, and
additionally may have significant self-absorption optical depth, and
so P$_R \propto \dot{m}_{\rm jet} t_{\rm inj} e^{-\tau}$ where
$\dot{m}_{\rm jet}$ is the (mean) rate of injection of relativistic electrons
into the jet, $t_{\rm inj}$ is the duration of this injection period,
and $\tau$ is the optical depth to self-absorption. Typically
$\tau_{\rm 5GHz} \leq 1$ by the peak or declining phases of the radio
events (see e.g.  Hjellming \& Han 1995), and so we do not consider
the $e^{-\tau}$ term 
to be significant at the level of accuracy of this paper.

\begin{figure}
\leavevmode\epsfig{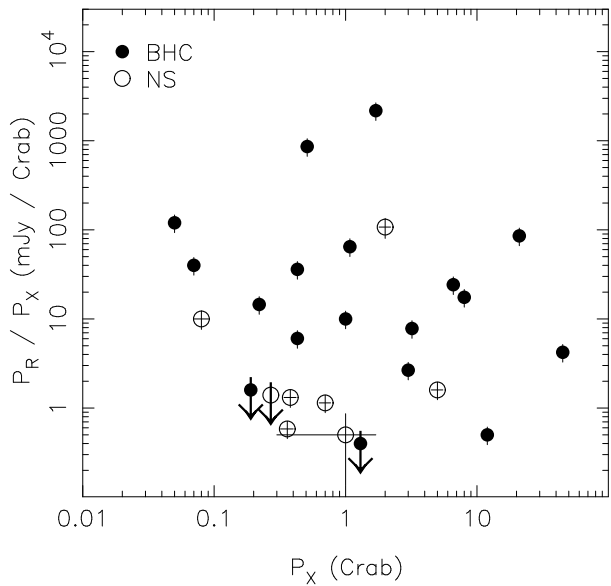}
\caption{A comparison of `radio loudness', P$_R$/P$_X$, against X-ray
peak flux for the NS and BHC systems.}
\end{figure}

\section{Correlations}

The two data sets presented in Table 1 and Fig 2 allow us to ask two
important questions, namely (a) are peak X-ray flux and peak radio
flux density correlated ?, and (b) is the relation between X-ray and
radio fluxes the same for the BHC and NS samples ? 

\subsection{Are P$_R$ and P$_X$ correlated ?}

We have tested the NS and BHC data sets individually using a
Spearman-Rank correlation test, and additionally for the entire
(NS+BHC+FT) data sets. The Spearman-Rank correlation coefficient for
each of the three samples, plus its significance, are presented in
table 2. Both the BHCs as an individual sample, and the entire
(NS+BHC+FT) sample show a statistically significant (nonparametric)
correlation at greater than the 95\% confidence level. While for the
NS sample the significance is only marginal (especially if 1A 1742-289
is excluded), this is
may be due to the small sample size. Therefore we can conclude that
the magnitude of X-ray and radio outbursts from X-ray transients are
probably
correlated.  As caveats to this, we note that this is still small
number statistics (particularly in the case of the NS systems), and
that we do not attempt to include the upper limits to the radio fluxes
reported for the 1998 Jan outburst of the Rapid Burster, the 1995 Aug
outburst of GRO J1655-40 and for XTE J1755-324.

\begin{table}
\centering
\caption{Spearman Rank-order correlation coefficient and associated
significance
for the neutron star, black hole candidate and combined (NS+BHC+FT)
groups}
\begin{tabular}{rccl}
\hline
Sample & N & $r_s$ & Significance \\
\hline
Neutron Star systems & 7 & +0.60 & $\sim 85$\% \\
NS without 1A 1742-289 & 6 & +0.41 & $\sim 60$\% \\
Black Hole systems & 16 & +0.59 & $\geq 97$\% \\
Combined (inc FTs) & 25 & +0.62 & $\geq 99$\% \\
\hline
\end{tabular}
\end{table}

\begin{figure}
\leavevmode\epsfig{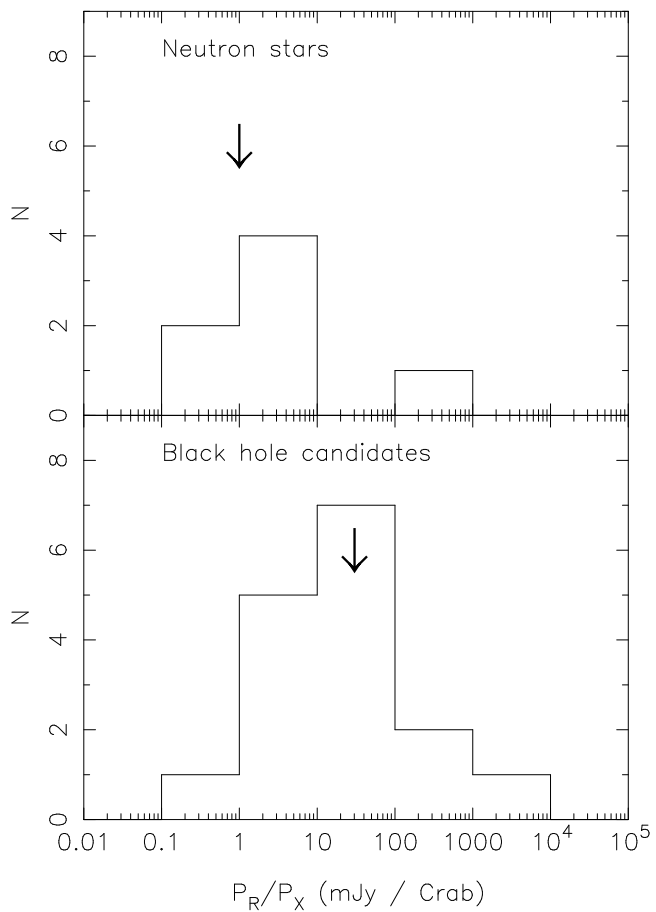}
\caption{Histograms of P$_R$/P$_X$, in order of magnitude bins, for
the NS and BHC samples. The BHC samples is clearly shifted to higher
values with respect to the NS sample; a K-S test formally confirms
that the two distributions are different (see text). The one NS datum
in the 100--1000 bin is associated with the source 1A 1742-289,
possibly the least reliable of the NS candidates (see text).
The arrows represent the
mean values for the persistent NS and BHC sources discussed in
Fender \& Hendry (2000) based on the radio data in that paper and
public XTE ASM data.
}
\end{figure}

\subsection{Black holes vs. Neutron stars}

We can quantitatively compare the NS and BHC data sets presented in
Table 1 and Fig 2 to see if we can discern any underlying differences
in the relation between X-ray and radio fluxes, and therefore
ultimately accretion and ejection, between the two types of compact
object. Firstly, we can apply a 2D Kolmogorov-Smirnov (K-S) test to
the two data sets to see if they are drawn from the same distribution
in P$_R$ vs. P$_X$.  The test reveals it is highly likely that the two
distributions are different (at a $\geq 97$\% confidence
level). However, this may simply reflect the fact that the NS sample
is from the lower end of the distribution. In Fig 3 we plot, for both
NS and BHC samples, `radio loudness' (P$_R$/P$_X$) as a function of
peak soft X-ray luminosity. Again a 2D K-S test reveals the two
samples, represented in this way, to be significantly
different (at $>98$\% confidence). 
Inspection of this Figure suggests that the discrepancy is
the distribution in (P$_R$/P$_X$), not in P$_X$. In Fig 4 we plot the
distribution of (P$_R$/P$_X$), in order of magnitude bins, for the NS
and BHC samples, reinforcing this impression. In addition in Fig 4 we
indicate the mean values of (P$_R$/P$_X$) for the six persistent
neutron star (specifically, Z-sources on the `Horizontal branch') and
four persistent black hole candidate (specifically, in the `Low/Hard'
X-ray state) binaries based on the radio data in Fender \& Hendry (2000) and
public RXTE ASM data.  A 1D K-S test confirms that the distributions in
(P$_R$/P$_X$) are significantly different ($>98$\% confidence)
and, interestingly, shows
that the distribution of P$_X$ is {\em not} significantly different
for the two samples ($<35$\% probability of null hypothesis).

Excluding upper limits, we find that for the NS sample, the mean value
($\pm$ standard error) of P$_R$/P$_X = 160
\pm 160$. Excluding the potentially anomalous and least reliable
source, ie. 1A 1742-289 (see section 2.2), reduces this substantially
to $2.6 \pm 1.5$. For the BHCs, we find P$_R$/P$_X = 230 \pm 150$
excluding the FTs, and P$_R$/P$_X = 230 \pm 130$ including them. Thus
this measure is consistent with BHC transients being more `radio-loud'
than NS transients, by a factor of between 1.5 -- 100.

\section{Discussion}

\subsection{Coupled inflow and outflow in transient events}

As discussed in section 3.1, P$_R$ and P$_X$ are significantly
positively correlated for the BHC X-ray transients (and marginally so
for the small NS sample). Based on section 2.3, this implies that
there is a correlation between $\dot{m}_{\rm disc}$ and $\dot{m}_{\rm
jet} t_{\rm inj} e^{-\tau}$. Since we argue in 2.3 that $\tau \la 1$
at the peak of the radio emission, we do not consider the $e^{-\tau}$
term significant for the order-of-magnitude arguments being presented
here. Therefore we can state that $\dot{m}_{\rm disc}$ is correlated
with $\dot{m}_{\rm jet} t_{\rm inj}$. We stress that the correlation
does not imply a simple linear relation, which is not a good fit to
either NS or BHC data sets. There is at least one physical reason
underlying this, namely that the radio emission, unlike the X-rays, is
expected to be beamed. If we could assume that $t_{\rm inj}$ is likely
to be related to the X-ray rise time of the outburst, and further that
$t_{\rm inj}$ is likely to be similar for all the systems, then we
could state confidently that $\dot{m}_{\rm disc}$ is correlated with
$\dot{m}_{\rm jet}$. However, Chen et al. (1997) find a fairly broad
distribution of rise times, generally of order one day, whose relation
to the radio outbursts we will not investigate here.  Nevertheless,
this is further evidence, not unexpectedly, that there is a clear
coupling between inflow and outflow around accreting compact objects.

\subsection{A higher radio:X-ray ratio for BHCs ?}

We have shown in section 3.2 that the BHC systems have, in general, a
higher ratio of P$_R$/P$_X$.  This result is consistent with the
common radio luminosity found for NS Z sources on the radio-bright
horizontal branch of the X-ray colour-colour diagram, and persistent
BHCs in the jet-producing Low/Hard X-ray state (Fender \& Hendry
2000), because the Z sources are generally around an order of
magnitude brighter in X-rays than the BHCs in the Low/Hard
state. Quantitatively, based on the mean radio flux densities
presented in Fender \& Hendry (2000) and public RXTE ASM data,
S$_R$/S$_X \sim 30$ for the persistent BHCs and S$_R$/S$_X \sim 1$ for
the persistent neutron stars (where S$_R$ and S$_X$ represent steady
fluxes for these systems); these values are indicated by arrows in Fig
4.  A word of caution: while it is reasonable to note qualitatively
the greater `radio loudness' of BHC systems in both tran sient and
persistent sources, a quantitative comparison of P$_R$/P$_X$ for the
transients with S$_R$/S$_X$ for the persistent sources is not
justified, as in the latter case there is likely to be significant and
continuous self-absorption at radio wavelengths (ie. $\tau_{\rm 5 GHz}
> 1$; Fender 2000c).

What is the underlying cause of this effect ? One possibility is that
for some reason black holes are more efficient at producing radio
emission per unit mass of accreted matter. Perhaps this is simply
because the deeper potential well of the black hole allows more energy
to be extracted, which beyond a certain X-ray flux is channelled into
the outflow, or maybe the black hole spin energy can be extracted via
the Blandford \& Znajek (1977) process. Another possibility is that
both classes of systems produce similar outflows at similar rates, but
that extra photons from the neutron star surface produce additional
cooling of the putative synchrotron-emitting electrons.  Finally, and
perhaps most likely, it may be not that black holes are `radio-loud',
but rather that they are `X-ray quiet', again related to the presence
of the event horizon rather than a surface. In this case we might
imagine the accretion flow would produce a jet at some radius which
was not affected by the nature of the accretor, but the flow was
radiatively inefficient and advected a large fraction of its energy
across the black hole event horizon 
(see e.g. Narayan, Mahadevan \& Quataert 1998).
It is interesting to note that in
section 3.2 we found that the distribution of P$_X$ was {\em not}
significantly different for the NS and BHC samples; this would appear
to support an `X-ray quiet' rather than a `radio loud' description of
the BHC sample, presuming that the BHCs are on average more massive
than the NS systems. Further comparisons of the distribution of P$_X$
for NS and BHCs, for {\em all} transient events, would be very
interesting (but would also require more careful consideration of
e.g. absorption and other selection effects).

It is interesting in this context to note the recent debate about the
relative luminosities of BHC and NS systems in quiescence, and their
ratio of outburst to quiescent X-ray flux (e.g. Asai et al. 1998;
Menou et al. 1999; Campana \& Stella 2000). While still
controversial, it has been suggested that the combination of advective
flow and event horizon results in a lower X-ray luminosity for
quiescent BHCs compared to quiescent NS systems. The results here,
while requiring further careful interpretation, {\em may} be further
evidence for broad observational differences between BHC and NS
systems resulting from the presence or not of a solid surface.

\subsection{Comparison with recurrent relativistic jet sources}

In Fig 5 we plot `radio loudness' as a function of P$_X$ for all the
systems in Table 1, including the range of peak fluxes displayed in
the repeated outbursts of Cyg X-3, GRS 1915+105 and Cir X-1 (based
somewhat subjectively on inspection of extended X-ray and radio data
sets). In addition we indicate the direction of a shift to higher peak
radio flux densities at smaller inclinations, if the radio emission is
significantly Doppler boosted.

Several things are noteworthy upon careful inspection of the figure:
\begin{itemize}
\item{The FTs are consistent with the distribution, in particular with the brightest of the BHCs}
\item{The repeated ($>100$; e.g. Pooley \& Fender 1997)
X-ray/radio outbursts of GRS 1915+105, frequently resolved into relativistic ejections (Mirabel \& Rodriguez 1994; Fender et al. 1999a), sit right in the middle of distribution of black hole outbursts}
\item{Cyg X-3 has a considerably higher radio:X-ray ratio than any of the 
other outbursting or transient systems}
\end{itemize}

We briefly discuss these points below :

\subsubsection{Fast transients}

The position of the FTs in Figs 2 and 3 reveals that their radio and X-ray
peak fluxes are entirely consistent with bright outbursts of BHC
systems. This is consistent with the results of Orosz et al. (2000a,b)
who claim that the FT V4641 Sgr is likely to contain a black hole (see
also the suggestion in Belloni et al. 1999 that the FT CI Cam may
contain a black hole).  However, given the small samples and strange
behaviour in the past of the e.g. NS Cir X-1, this cannot be considered as
more than circumstantial support for this interpretation.

\subsubsection{Cyg X-3} 

Cyg X-3 has the largest radio:X-ray ratio of {\em any} X-ray binary;
in fact with multiple radio outbursts brighter than 10 Jy
(e.g. Waltman et al. 1995), it is the brightest radio source ever
associated with an X-ray binary. Furthermore, it was one of the
earliest sources in which a clear correlation between X-ray and radio
outbursts was established (Watanabe et al. 1994; see also McCollough
et al. 1999).  Why is it so bright ? Recent VLBA observations reveal a
milliarcsecond-scale jet which is clearly one-sided (Mioduszewski et
al.  1998, 2001) and may be interpreted as a $\ga 0.8c$ jet inclined within
a few degrees of the line of sight. If so, Cyg X-3 is the galactic 
equivalent of a blazar and may represent more or less the extreme 
of the possible radio:X-ray ratios.

\begin{figure}
\leavevmode\epsfig{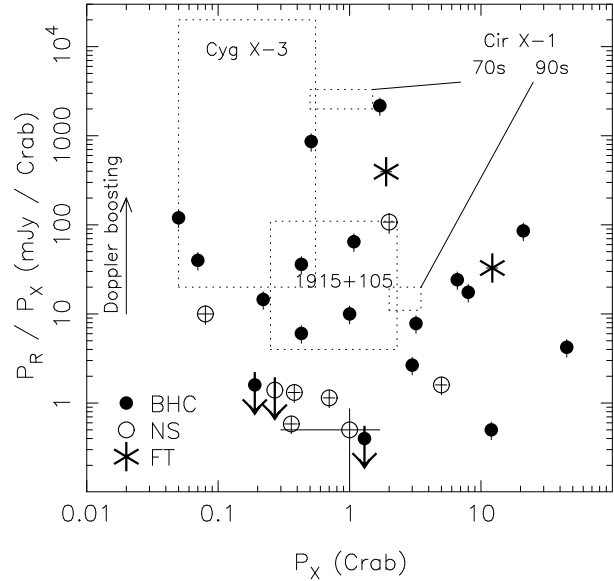}
\caption{A composite figure displaying the data for both neutron star and black hole candidate systems together, as well as indicating typical ranges of
 variability for the recurrently-outbursting systems Cyg X-3, GRS 1915+105 and
 Cir X-1 (the latter for epoch 1970s and epoch 1990s).
}
\end{figure}

\subsubsection{GRS 1915+105}

As Fig 5 clearly demonstrates, the repeated simultaneous X-ray:radio
outbursts of the `microquasar' GRS 1915+105 lie right in the middle of
the BHC distribution. This clearly indicates that while the repetition
rate and timescales of X-ray and radio behaviour exhibited by GRS
1915+105 may be unique (e.g. Belloni et al. 2000), the peak X-ray and
radio fluxes for individual outbursts are entirely `normal' for BHCs
in outburst.

\section{Conclusions}

In collecting together data on (near-)simultaneous peak X-ray and
radio fluxes from neutron-star (NS), black-hole-candidate (BHC) and a
small number of other systems, we have established that there
is a significant positive correlation between the two measures for the BHC
sample. The NS sample may also be correlated, but the small numbers at
present result in only marginal evidence.
This is further direct evidence of a
coupling between accretion and jet formation in X-ray binaries, and in
this case indicates qualitatively that the maximum disc accretion rate
is related to the total number of synchroton-emitting electrons
ejected.  In principle, the accretion and outflow rate could be
directly compared using peak X-ray flux (P$_X$) and rate of increase
of the radio flux density before the peak of the radio outburst
($dS_{\nu}/dt$), but since in most cases the radio rise is missed,
this will not be easy to achieve.

We have also demonstrated that the distribution of
P$_R$/P$_X$, or `radio loudness', is significantly different for the
NS and BHC distributions, in the sense that the BHC sample is in
general more `radio loud'. This is confirmed by comparison with the
persistent, bright BHC and NS systems. The root of this difference
seems likely, as with most observational differences between neutron
star and black hole X-ray binaries, to be in the presence of a surface
in only the NS sample. An important question is whether the BHC sample
is `radio loud' or `X-ray quiet'. The latter interpretation, if
confirmed, may be considered as further evidence for advective flows
crossing an event horizon in black hole systems (but presumably in
this case at high accretion rates). This result is
particularly interesting in the light of recent claims that luminosity
differences between quiescent BHS and NS systems are also tracers of
the presence or not of a solid surface (e.g. Asai et al. 1998; Menou
et al. 1999; Campana \& Stella 2000).

Furthermore, at least three
of the BHC sample and both of the FT sample have been
resolved into moderately relativistic outflows at radio wavelengths.
Their consistency with the rest of the sample implies that the
simplest picture is one in which {\em all} BHC transients produce a
mildly relativistic (bulk Lorentz factor $1 \leq \Gamma \leq 5$)
jet (if $\Gamma$ were very high, it would destroy the correlation and
would also result in more systems with no detectable radio emission,
as it would be beamed out of the line of sight).
The neutron stars probably also produce such outflows, but direct
observational measurements of $\Gamma$ for this sample do not yet
exist. An origin for all X-ray transient radio outbursts in outflows is
not a new idea -- see e.g. Hjellming \& Han (1995) -- but this is
the first quantitative investigation of the concept.
Supporting this picture, the range of peak fluxes associated
with the multiple X-ray/radio outbursts of the jet source GRS
1915+105, are entirely consistent with the broader distribution of BHC
transients.  This in turn implies that there is nothing special about
the physics of the accretion : outflow coupling in GRS 1915+105.
While we have not attempted to constrain the upper bound to $\Gamma$,
the reader should bear in mind that the higher its (mean) value, the
less likely it is that we should find the observed correlation, as a
result of beaming (relativistic aberration), {\em unless the X-rays
are also beamed}.  This point is discussed further in Fender \& Hendry
(2000) for the case of persistent X-ray binaries, where an attempt is
made to constrain the maximum value of $\Gamma$ using Monte Carlo
simulations.

Exceptions to this simplistic model of course exist; the rapid
transition from radio-loud to radio-quiet during a sequence of
declining X-ray outbursts of GRO J1655-40 (corresponding to a decrease
in P$_R$/P$_X$ by a factor of 5500!) is hard to explain.  Similarly,
the decay by orders of magnitude in the strength of radio outbursts
from Cir X-1 while its peak X-ray flux has increased by a factor of
2--3 is difficult to understand. It seems probable that more than
simply jet precession is required to explain some changes, and that
additional factors e.g. thermal absorption will need to be considered
before the picture is clear.

\section*{Acknowledgements}

This research has made use of the SIMBAD database, operated at CDS,
Strasbourg, France. We thank G. G. Pooley and the late
R. M. Hjellming and for continuous free exchange of information,
P. Jonker, J. Homan and R. Wijnands for discussions about our data
set, and both the anonymous referee and G. Dubus for critical reading of the
manuscript which led to significant improvements to the paper.

\end{document}